\documentclass{webofc}
\usepackage[varg]{txfonts}   % Web of Conferences font
\usepackage{listings}
\usepackage{hyperref}
\begin{document}
\title{Gaia's  Cepheids and RR Lyrae Stars and  Luminosity Calibrations\linebreak
Based on Tycho-Gaia Astrometric Solution}
%
% subtitle is optionnal
%
%%%\subtitle{Do you have a subtitle?\\ If so, write it here}

\author{\firstname{Gisella} \lastname{Clementini}\inst{1}\fnsep\thanks{\href{mailto:gisella.clementini@oabo.inaf.it}{\tt gisella.clementini@oabo.inaf.it}} \and
             \firstname{Laurent} \lastname{Eyer}\inst{2}  \and
             \firstname{Tatiana} \lastname{Muraveva}\inst{1} \and
             \firstname{Alessia} \lastname{Garofalo}\inst{1,3} \and
             \firstname{Vincenzo} \lastname{Ripepi}\inst{4} \and
             \firstname{Marcella} \lastname{Marconi}\inst{4} \and
             \firstname{Luis} \lastname{Sarro}\inst{5} \and
             \firstname{Max} \lastname{Palmer}\inst{6} \and
             \firstname{Xavier} \lastname{Luri}\inst{6} \and
             \firstname{Roberto} \lastname{Molinaro}\inst{4} \and
             \firstname{Lorenzo} \lastname{Rimoldini}\inst{7} \and
             \firstname{Laszlo} \lastname{Szabados}\inst{8} \and
             \firstname{Richard~I.} \lastname{Anderson}\inst{9} \and     
             \firstname{Ilaria} \lastname{Musella}\inst{4}
}

\institute{
         INAF - Osservatorio Astronomico di Bologna, Via Gobetti 93/3, 40129 Bologna, Italy 
\and
         Department of Astronomy, University of Geneva, Ch. des Maillettes 51, 1290 Versoix, Switzerland
\and
         Dipartimento di Fisica e Astronomia, Universit\`a di Bologna, Via Gobetti 93/2, 40129 Bologna, Italy
\and
        INAF-Osservatorio Astronomico di Capodimonte, Via Moiariello 16, 80131 Napoli, Italy
\and
        Dpto. Inteligencia Artificial, UNED, c/ Juan del Rosal 16, 28040 Madrid, Spain
\and
       Institut de Ci\`{e}ncies del Cosmos, Universitat  de  Barcelona  (IEEC-UB), Mart\'{i}  Franqu\`{e}s  1, E-08028 Barcelona, Spain
\and
       Department of Astronomy, University of Geneva, Ch. d'Ecogia 16, 1290 Versoix, Switzerland  
\and
      Konkoly Observatory, Research Centre for Astronomy and Earth Sciences, Hungarian Academy of Sciences, Konkoly Thege Mikl\'{o}s \'{u}t 15-17, 1121 Budapest, Hungary
\and
      Department of Physics and Astronomy, The Johns Hopkins University, 3400 N Charles St, Baltimore, MD 21218, USA
      }

\abstract{%% error
Gaia Data Release 1 contains 
parallaxes for  more than  700 Galactic Cepheids  and RR Lyrae stars, computed as   part of the Tycho-Gaia Astrometric Solution  (TGAS).   
We have used TGAS parallaxes, along with literature ($V, I, J, {K_\mathrm{s}}, W_1$) photometry and spectroscopy, to calibrate the zero point of  
 the Period-Luminosity and Period-Wesenheit  relations of classical and type
II Cepheids, and the near-infrared Period-Luminosity, Period-Luminosity-Metallicity and optical Luminosity-Metallicity 
relations of RR Lyrae stars. In this contribution we briefly summarise results obtained  by fitting these basic relations   
adopting different techniques that operate either in parallax or distance (absolute magnitude) space. 
 }
\maketitle

\section{Introduction}\label{sec:intro}
Cepheids and RR Lyrae stars are primary standard candles of  the cosmological distance ladder and excellent 
tracers of young (the classical Cepheids)  and old (the RR Lyrae stars and the Type~II Cepheids) stellar populations. Gaia will be an extraordinary 
discoverer of new Cepheids and RR Lyrae stars in and beyond  the Milky Way.  Furthermore, the unprecedented accuracy of 
Gaia's  end-of-mission astrometric measurements will allow  us to pin down  
slope and zero point of the fundamental relations that Cepheids  and RR Lyrae stars conform to 
 with the  precision  required to constrain the Hubble constant better than 2\%. 
 
Besides $G$-band time-series photometry and pulsation characteristics for a few thousand Cepheids and RR Lyrae stars in the 
Large Magellanic Cloud (LMC; \cite{RefCle16}, see also Eyer et al., this volume), data for variable stars in Gaia Data Release 1 (DR1) include parallaxes for 
331 classical Cepheids, 31 Type~II Cepheids and 364 RR Lyrae stars, 
in common between Gaia and the Hipparcos and Tycho-2 catalogues. They were computed as part of the Tycho-Gaia  Astrometric Solution (TGAS, \cite{RefLind16}).

\begin{figure}
\centering
\includegraphics[trim=30 300 0 350 clip, width=0.91\linewidth]{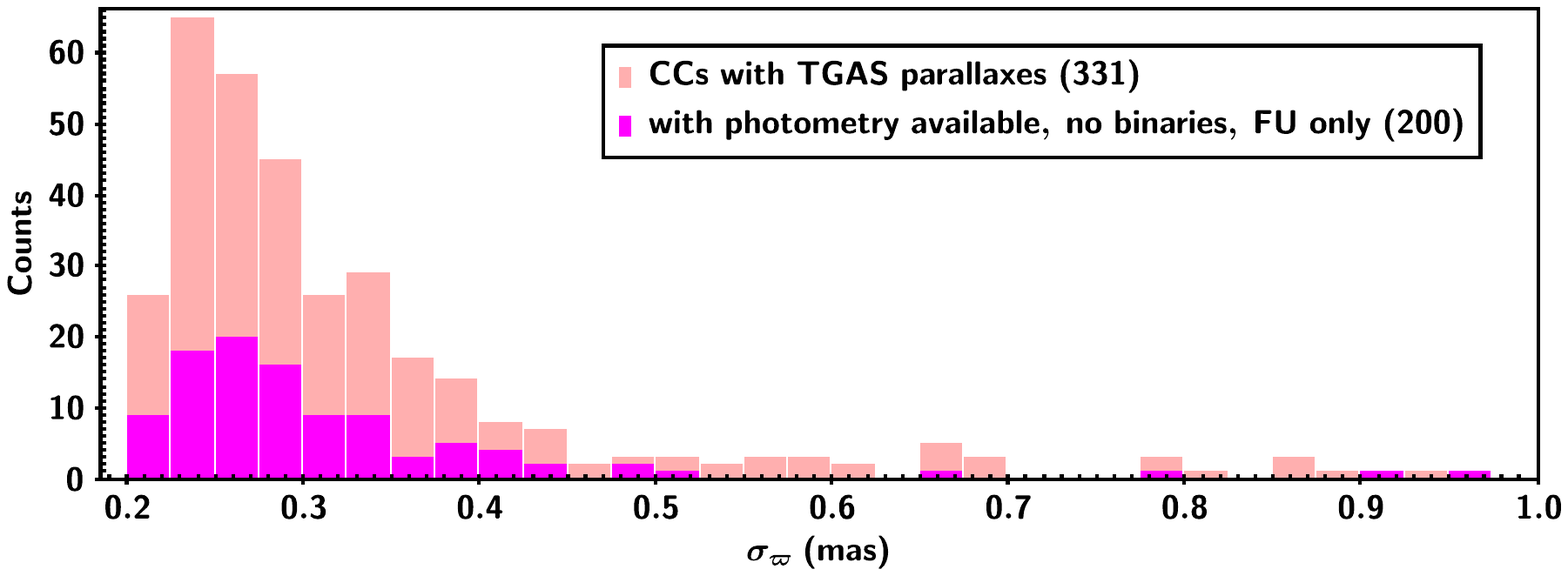}
\caption{Error distributions of TGAS parallaxes for classical Cepheids. The pink histogram corresponds to the whole sample of 331 Galactic classical Cepheids  in the TGAS catalogue,  the magenta histogram  shows a subsample of 102 fundamental mode classical Cepheids 
with ($V, I, J, K_\mathrm{s}$) photometry available in the literature after removing the known binaries. 
The bin size is 0.025~mas.}
 % Give a unique label
\label{fig1}       % Give a unique label
\end{figure}
\begin{figure}
\centering
\includegraphics[trim=30 300 0 310 clip, width=0.91\linewidth]{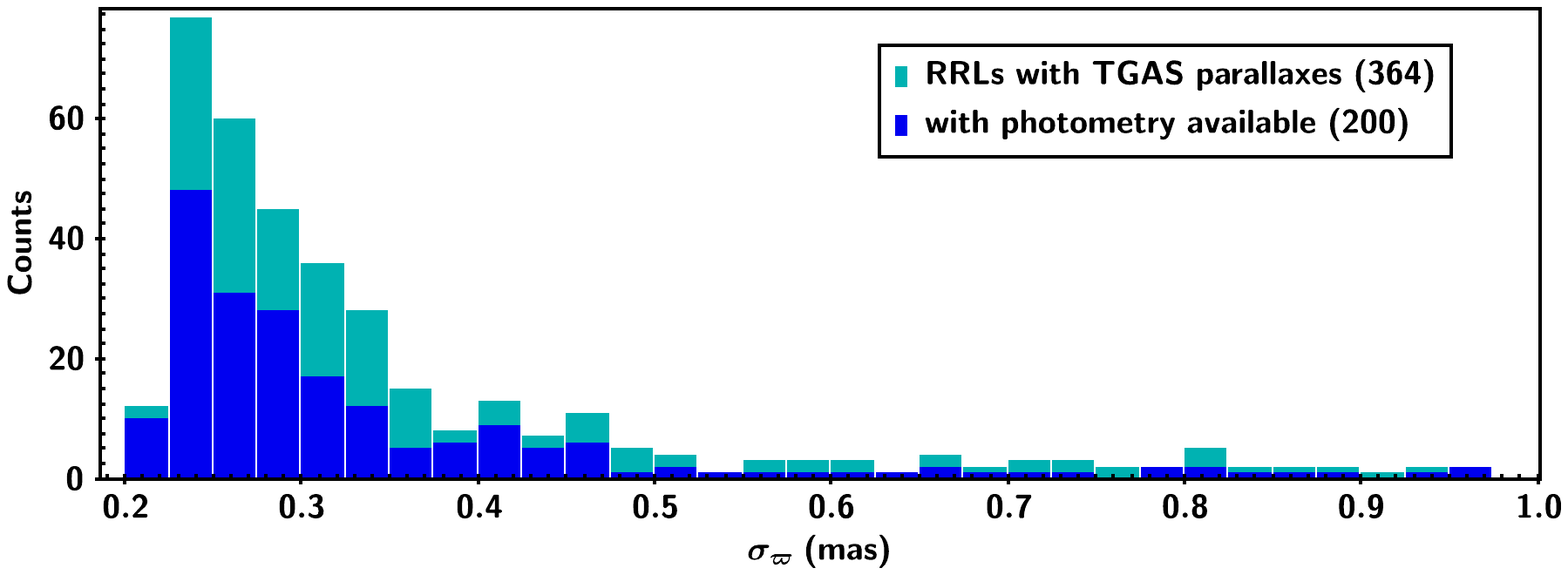}
\caption{Error distributions of TGAS parallaxes for RR Lyrae stars.  The cyan histogram corresponds to the whole sample  of 364 RR Lyrae stars  in the TGAS catalogue,  the blue histogram shows a 
subsample 
of 200 stars with ($V, K_{\rm s}, W_1$) photometry available in the literature. The bin size is 0.025~mas.}
\label{fig2}       % Give a unique label
\end{figure}

The error distributions of the TGAS parallaxes for classical Cepheids 
and RR Lyrae stars are shown in Figs.~\ref{fig1}  
and ~\ref{fig2}, respectively. Errors  
range from 0.2 to about 1 milliarcsecond (mas) and peak around  $\sim$ 0.25 mas for both types of pulsating stars\footnote{After publication of  the Gaia DR1 catalogue, a number of authors have suggested that the standard errors of TGAS parallaxes may be overestimated (e.g. \cite{RefCas17} and  Michael 
Feast's concluding comments, this volume).}. Although this  is not comparable to the final Gaia precision, it already represents a significant general improvement with respect to 
Hipparcos parallaxes (\cite{RefVan07}).  Additionally, as described in \cite{RefLind16},  there could still be some 
systematic
    effects at a typical level of $\pm$ 0.3~mas depending on the sky position and the colour of the source.
\begin{figure}
\centering
\sidecaption
% Use the relevant command for your figure-insertion program
% to insert the figure file. See example above.
% If not, use
\includegraphics[trim=20 160 30 100 clip, width=6.5cm,clip]{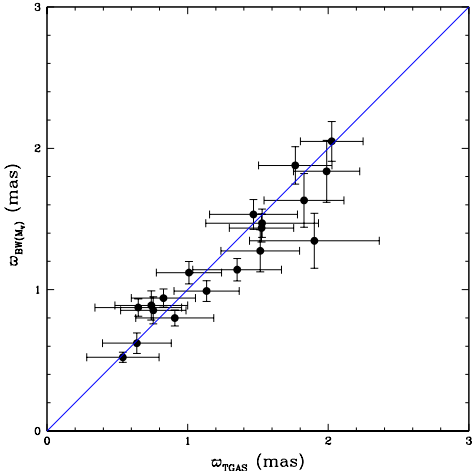}
\caption{Comparison between the TGAS parallaxes and  photometric parallaxes inferred from the absolute visual magnitude ($M_V$) estimated via the Baade-Wesselink (B-W)  technique for 19 Galactic RR Lyrae stars (see table 2 in \cite{RefMura15} and references therein). The blue line represents the bisector.}
\label{fig3}       % Give a unique label
\end{figure}
However, these systematic effects are spatially correlated and 
become negligible for RR Lyrae stars and Type~II Cepheids  
that are evenly distributed on the sky, while may be relevant for classical Cepheids, which are mainly  concentrated in  the Galactic disk. 
We also note that the TGAS samples are the result
of several processing stages  and truncations (e.g., removal of sources brighter than a certain limit and bluer/redder than a certain colour interval, truncation of sources with less than 5 transits,  or with large astrometric uncertainties, see \cite{RefCle17} for details), each with a different impact on
the resulting sample. Nevertheless, the very limited size of the Cepheid and RR Lyrae samples and the large relative errors of TGAS parallaxes hide most of these effects.

\section{Comparison with trigonometric and photometric parallaxes in the literature}\label{sec:samples}
A direct comparison between TGAS and Hipparcos
 parallaxes (\cite{RefVan07}) is possible for 248 classical
Cepheids, 31 Type II Cepheids and 188 RR Lyrae stars for which both measurements are available. 
  The number of negative parallaxes is significantly reduced from 32\% of the sample of classical Cepheids in Hipparcos
to  only 4\% in TGAS, decreasing from 42\% to 16\% for the Type~II Cepheids and  from 32\% to 1\% for the RR Lyrae stars. This is a clear indication that uncertainties of the TGAS parallaxes are smaller than in Hipparcos.  When comparing TGAS and Hipparcos parallaxes  
in the Period-Luminosity in the $K_{\mathrm{s}}$-band ($PL_{K_\mathrm{s}}$) plane, the improvement  in quality and statistics of the former is impressive, particularly  for RR Lyrae stars. 

A direct comparison of TGAS versus  HST is possible for three classical Cepheids with HST parallaxes from \cite{RefBen07},\cite{RefRie14} and  \cite{RefCas16}, one Type~II Cepheid (VY Pyx) 
 and five RR Lyrae stars with HST parallaxes from \cite{RefBen11}. TGAS and HST parallaxes compare favourably for the classical Cepheids and even better for the RR Lyrae stars. On the other hand,  the TGAS parallax  of  VY Pyx  is much smaller  than the HST value by \cite{RefBen11}, and nicely places the star  on the
extrapolation to longer periods of the RR Lyrae star $PL_K$ relation (see figure 10, in \cite{RefCle17}), while this is not the case with the HST parallax. 

Figure~\ref{fig3} shows the comparison between the TGAS parallaxes and  photometric parallaxes inferred from the absolute visual magnitude ($M_V$) estimated via the Baade-Wesselink (B-W)  technique for 19 Galactic RR Lyrae stars (see table
2 in \cite{RefMura15} and references therein). The agreement is excellent both in the case of the $M_V$ and the $K$-band absolute magnitudes ($M_K$). 
Similarly, good agreement is found between TGAS parallaxes and the photometric parallaxes inferred from the application of the Infrared Surface
Brightness version of the B-W technique to 54 classical
Cepheids from the collection in \cite{RefFou07}. 
Pulsation parallaxes for a few classical Cepheids and RR Lyrae stars in our samples have been estimated from the theoretical modelling
of the star multi-band light curves through nonlinear convective pulsation models (Marconi, this volume) and compared with the corresponding 
TGAS parallaxes. Agreement  is generally satisfactory  except for RS Cas, a classical Cepheid for which TGAS parallax is likely incorrect as it places the star about 2 magnitudes below the $PL$ relation. 

\section{Luminosity calibrations based on different fitting approaches}
A procedure often used to calibrate the Period-Luminosity ($PL$), Period-Wesenheit ($PW$), Period-Luminosity-Metallicity ($PLZ$) or Luminosity-Metallicity ($M_V$ - [Fe/H]) relations is the direct transformation to distance (hence, absolute magnitude)
by parallax inversion and then the least squares fit  (LSQ) of
the derived parameters. However, symmetrical errors in parallaxes translate into asymmetric 
errors in the magnitudes, an effect that becomes specially problematic for 
parallaxes with  large relative errors that introduces an overall bias. Furthermore, 
this method  does not allow us to use negative parallaxes, thus biasing the samples 
against more distant sources.
 On the contrary, in methods that operate
in parallax space such as the Astrometric Based Luminosity
(ABL, \cite{RefAre99}) and Bayesian approaches (hereinafter, BA)
the parallaxes are used
directly, thus maintaining symmetrical the errors and allowing negative parallaxes to be used. We applied all three methods (LSQ, ABL and BA) to fit the canonical relations that Cepheids and RR
Lyrae stars conform to. In Table~\ref{tab:tab-1}, we summarise  the TGAS-based $PL_{K_\mathrm{s}}$ relations of Cepheids and RR Lyrae stars obtained with the three different fitting 
approaches and adopting the slope from  \cite{RefFou07} for classical Cepheids, from \cite{RefRip15} for the Type~II Cepheids and from \cite{RefMura15} for the RR Lyrae stars. 
\begin{table}
\centering
\caption{$PL_{K_{\rm s}}$ relations for classical Cepheids, Type~II Cepheids and RR Lyrae stars with zero point based on TGAS parallaxes.}
\begin{tabular}{lcc}
\hline
 & Relation (mag) & r.m.s. (mag) \\\hline
\noalign{\smallskip}
$PL_{K_\mathrm{s}}$ 95 objects (LSQ) & $-3.365\log P-(2.06\pm0.08)$ & 0.74  \\
$PL_{K_\mathrm{s}}$ 102 objects  (ABL) & $-3.365\log P-(2.63\pm0.10)$ & 0.88 \\
$PL_{K_\mathrm{s}}$ 102 stars (BA) & $ -3.365 \log P - (2.60\substack{+0.11\\-0.15} ) $& 1.33  \\\hline
%\noalign{\smallskip}
$PL_{K_\mathrm{s}}$ 22 objects (LSQ) & $-2.385\log P-(1.18\pm0.12)$ & 0.81  \\
$PL_{K_\mathrm{s}}$ 26 objects  (ABL) & $-2.385\log P-(1.58\pm0.17)$ & 1.10  \\
$PL_{K_\mathrm{s}}$ 26 objects (BA) & $-2.385 \log P - (1.51\substack{+0.23\\-0.22} ) $& 1.14  \\\hline
%\noalign{\smallskip}
$PL_{K_\mathrm{s}}$ 195 stars (LSQ) & $-2.73\log P - (1.06 \pm 0.04)$ & 0.84  \\
$PL_{K_\mathrm{s}}$ 200  stars (ABL) & $-2.73\log P - (1.26 \pm 0.04)$ & 0.90  \\
$PL_{K_\mathrm{s}}$ 200  stars (BA) & $-2.73\log P - (1.24 \pm 0.05)$ & 1.02  \\\hline 
%\noalign{\smallskip}
\end{tabular}
\label{tab:tab-1}       % Give a unique label
\end{table}
Table~\ref{tab:tab-1} shows
that the ABL and Bayesian approaches are generally in good
agreement with  each other and provide brighter absolute magnitudes
(hence longer distances) than the direct transformation
of parallaxes and the LSQ fit. 
Differences are larger (of about 0.5-0.6~mag, on average) for the classical Cepheids, reduce to 0.4-0.5 mag for the
Type II Cepheids, and are the smallest ones, 0.2~mag, for the RR
Lyrae stars. However, we 
note that the r.m.s. scatter of all relations is very large, due
to the large parallax uncertainties. 
\begin{figure}
\centering
\sidecaption
\includegraphics[trim= 20 175 190 240 clip, width=8.0cm]{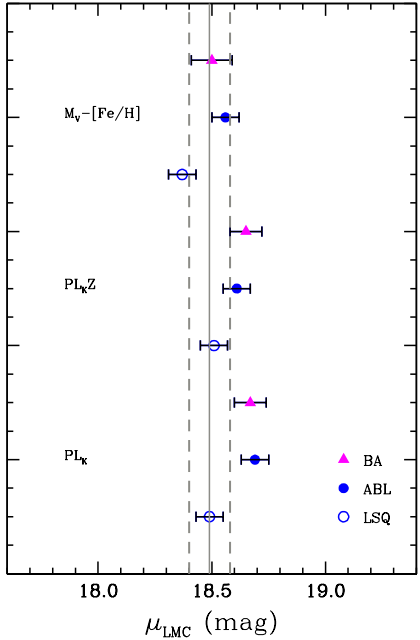}
\caption{LMC distance moduli obtained by fitting the TGAS-based  $PL_K$, $PL_KZ$ and $M_{V}-{\rm [Fe/H]}$  relations of RR Lyrae stars with the LSQ (blue open circle),  the ABL  (blue filled circles) and the Bayesian approaches (magenta filled triangles), respectively.
From  bottom to top: $PL_{K_{\rm s}}$ relations for 195/200/200 stars (LSQ, ABL, BA methods)  with 
slope from
\cite{RefMura15}; $PL_{K_{\rm s}}Z$ relations 
with slope of the dependence on period from
\cite{RefMura15}; $M_{V}-{\rm [Fe/H]}$ relations 
with slope from \cite{RefCle03}. A solid vertical line shows  the LMC distance modulus from \cite{RefPie13}.}
\label{fig4}
\end{figure}

Figure~\ref{fig4} shows the results obtained for the LMC distance modulus 
by applying the TGAS-based relations of the RR Lyrae stars to the LMC variable stars 
studied by \cite{RefCle03}. The results obtained for the RR
Lyrae stars show a much better agreement among the three
methods and also a reasonably good agreement with the currently adopted LMC distance modulus
from  \cite{RefPie13} (solid, vertical line).

The TGAS parallaxes represent a significant improvement  upon the previous Hipparcos estimates. 
However, the TGAS-based luminosity calibrations presented in this study have to be considered preliminary and  to 
 be superseded by new more accurate relations calibrated on Gaia-only parallaxes that will be published in future releases, to start with Gaia Data Release 2 in 2018.

\begin{acknowledgement} 
\noindent\vskip 0.2cm
\noindent {\em Acknowledgments}
This work has made use of data from the ESA space mission Gaia, processed by the Gaia Data Processing and Analysis Consortium (DPAC) and of % This research has made use of 
the SIMBAD database, operated at CDS, Strasbourg, France.
\end{acknowledgement}


\begin{thebibliography}{}

% Format for Conference Proceedings paper 
\bibitem{RefAre99}
F. Arenou, X. Luri,  in {\it Harmonizing
Cosmic Distance Scales in a Post-Hipparcos Era}, ASP Conference Series, Vol. 167, ed. D. Egret \& A. Heck, p. 13 (1999)
%astro-ph/9812094

\bibitem{RefBen07}
Benedict, G.F., McArthur, B.E., Feast, M.W., et al., AJ, 133, 1810 (2007)

\bibitem{RefBen11}
Benedict, G.F., McArthur, B.E., Feast, M.W., et al., AJ, 142,187 (2011)

\bibitem{RefCas16}
Casertano, S., Riess, A.G., Anderson, J., et al., ApJ, 825, 11 (2016)

\bibitem{RefCas17}
Casertano, S., Riess, A.G., Bucciarelli, B., \& Lattanzi, M.G., A\&A, 599, A67 (2017)

\bibitem{RefCle03}
Clementini, G., Gratton, R.G., Bragaglia, A., et al., AJ, 125, 1309 (2003)

\bibitem{RefCle17}
Gaia Collaboration, G. Clementini, et al.,  A\&A, in press (2017)

\bibitem{RefCle16}
Clementini, G., Ripepi, V. Leccia, S., 
et al., A\&A 595, A133 (2016)%, 1609.04269

\bibitem{RefFou07}
Fouqu\'e, P., Arriagada, P., Storm, J., et al., A\&A, 476, 73 (2007)

\bibitem{RefLind16}
Lindegren, L., Lammers, U., Bastian, U., et al., A\&A, 595, A4 (2016)

\bibitem{RefMura15}
Muraveva, T.,  Palmer, M.,  Clementini, G., et al.,
ApJ, 807, 127 (2015)
% Format for books

\bibitem{RefPie13}
Pietrzynski, G., Graczyk, D., Gieren, W., et al., Nature, 495, 76 (2013)

\bibitem{RefRie14}
Riess, A.G., Casertano, S., Anderson, J., MacKenty, J., Filippenko, A.V., ApJ, 785, 161 (2014)

\bibitem{RefRip15}
Ripepi, V., Moretti, M.-I., Marconi, M., et al., MNRAS, 446, 3034 (2015)

\bibitem{RefVan07}
van Leeuwen, F.,  A\&A, 474, 653 (2007)

\end{thebibliography}
\end{document}